\documentclass[a4paper,fleqn,usenatbib]{mnras}
\usepackage{Times}
\usepackage[T1]{fontenc}
\usepackage{ae,aecompl}
\usepackage{graphicx}	
\usepackage{amsmath}	
\usepackage{amssymb}	

\title[Environment and SFGs at $z\sim0.4$]{The nature of H$\alpha$ star-forming galaxies at $\bf z\sim0.4$ in and around Cl 0939+4713: the environment matters\thanks{Based on observations obtained with AF2+WYFFOS on the WHT, programme W14BN020.}}
\author[D. Sobral et al.]{David Sobral$^{1,2,3,4}$\thanks{VENI Fellow. E-mail: d.sobral@lancaster.ac.uk}, Andra Stroe$^{5,2}$\thanks{ESO fellow}, Yusei Koyama$^{6}$, Behnam Darvish$^{7}$, \newauthor Jo{\~a}o Calhau$^{1,3,4}$, Ana Afonso$^{3,4}$, Tadayuki Kodama$^{8,9}$, Fumiaki Nakata$^{6}$ \\
$^{1}$ Department of Physics, Lancaster University, Lancaster, LA1 4YB, UK \\
$^{2}$ Leiden Observatory, Leiden University, P.O.\ Box 9513, NL-2300 RA Leiden, The Netherlands \\
$^{3}$ Instituto de Astrof\'{\i}sica e Ci\^{e}ncias do Espa\c{c}o, Universidade de Lisboa, OAL, Tapada da Ajuda, PT1349-018 Lisbon, Portugal \\
$^{4}$ Departamento de F\'{i}sica, Faculdade de Ci\^{e}ncias, Universidade de Lisboa, Edif\'{i}cio C8, Campo Grande, PT1749-016 Lisbon, Portugal \\
$^{5}$ European Southern Observatory, Karl-Schwarzschild-Str. 2, 85748, Garching, Germany  \\
$^{6}$ Subaru Telescope, National Astronomical Observatory of Japan, 650 North A'ohoku Place, Hilo, HI 96720, U.S.A. \\
$^{7}$ Cahill Center for Astrophysics, California Institute of Technology, 1216 East California Boulevard, Pasadena, CA 91125, USA \\
$^{8}$ Department of Astronomical Science, The Graduate University for Advanced Studies (SOKENDAI), Mitaka, Tokyo 181-8588, Japan \\
$^{9}$ Optical and Infrared Astronomy Division, National Astronomical Observatory of Japan, Mitaka, Tokyo 181-8588, Japan}
\date{Accepted 2016 March 1. Received 2016 March 1; in original form 2015 December 5}
\pubyear{2016}

\begin{document}
\label{firstpage}
\pagerange{\pageref{firstpage}--\pageref{lastpage}}
\maketitle

\begin{abstract}  
Cluster star-forming galaxies are found to have an excess of Far-Infrared emission relative to H$\alpha$, when compared to those in the field, which could be caused by intense AGN activity, dust and/or declining star formation histories. Here we present spectroscopic observations of H$\alpha$ emitters in the Cl 0939+4713 (Abell 851) super-cluster at $z=0.41$, using AF2+\,WYFFOS on the WHT. We measure [O{\sc ii}], H$\beta$, [O{\sc iii}], H$\alpha$ and [N{\sc ii}] for a sample of 119 H$\alpha$ emitters in and around the cluster. We find that $17\pm5$\% of the H$\alpha$ emitters are AGN, irrespective of environment. For star-forming galaxies, we obtain Balmer decrements, metallicities and ionisation parameters with different methods, individually and by stacking. We find a strong mass-metallicity relation at all environments, with no significant dependence on environment. The ionisation parameter declines with increasing stellar mass for low-mass galaxies. H$\alpha$ emitters residing in intermediate environments show the highest ionisation parameters (along with high [O{\sc iii}]/H$\alpha$ and high [O{\sc iii}]/[O{\sc ii}] line ratios, typically twice as large as in the highest and lowest densities), which decline with increasing environmental density. Dust extinction (A$_{\rm H\alpha}$) correlates strongly with stellar mass, but also with environmental density. Star-forming galaxies in the densest environments are found to be significantly dustier (A$_{\rm H\alpha}\approx1.5-1.6$) than those residing in the lowest density environments (A$_{\rm H\alpha}\approx0.6$), deviating significantly from what would be predicted given their stellar masses.
\end{abstract}


\section{Introduction}

The last decades have seen a tremendous observational, modelling and theoretical effort towards understanding galaxy formation and evolution in various environments. One of the key observables that has now been widely measured is the star formation rate density ($\rho_{\rm SFR}$), which many studies have shown to rise out to $z\sim2$ \citep[e.g.][]{Lilly96,Karim11,Burgarella13,Sobral13}. Measuring $\rho_{\rm SFR}$ as a function of redshift ($z$), but also understanding its variation with environment and with internal processes is key to our understanding and to test state-of-the-art models \citep[e.g. Illustris, EAGLE;][]{Vogelsberger2014,Genel2014,Schaye2015,Crain2015C}.

In the local Universe, star formation depends strongly on environment. Clusters are primarily populated by passive galaxies, while star-forming galaxies are mainly found in lower-density environments \citep[e.g.][]{Dressler1980}. It is also well-established \citep[e.g.][]{Best04} that the fraction of star-forming galaxies decreases with increasing local galaxy density (often projected local density, $\Sigma$) both in the local Universe and at low ($z<0.5$) redshift \citep[e.g.][]{Kodama2004}. Stellar mass and/or internal processes connected with the stellar mass assembly also play a key role. While mass and environmental density correlate, it is now possible to disentangle their roles in the local Universe and at higher redshift and to show that both are relevant for quenching star-formation \citep[e.g.][]{Peng10,Darvish16}.

Several studies in a broad range of redshifts have shown that, on average, many properties of star-forming galaxies that are directly or indirectly linked to star formation activity (e.g. star formation rates, specific star formation rates, emission line equivalent widths, main-sequence of star-forming galaxies) seem to be invariant to their environment \citep[e.g.][]{Peng10,Wijesinghe12,Muzzin12,Koyama13,Koyama14,Hayashi14,Darvish14,Darvish15,Darvish16}. Therefore, the main role of the environment seems to be to set the fraction of quiescent/star-forming galaxies \citep[e.g.][]{Peng10,Sobral11,Muzzin12,Darvish14,Darvish16}. Nevertheless, recent studies are finding that not all characteristics of star-forming galaxies are independent of environment. For example, metallicities and electron densities have been shown to be a function of environment \citep[e.g.][]{Kulas2013,Shimakawa15}, with studies finding that star-forming galaxies have much lower electron densities and slightly higher metallicities in high density environments when compared to lower density/more typical environments \citep[][]{Sobral2015,Darvish15}.

Finding the exact mechanisms of galaxy quenching and their physical agents is still one of the unsolved problems in galaxy evolution. Many internal (e.g. stellar and AGN feedback) and external (e.g. galaxy environment) physical drivers are thought to be linked to the quenching process. Several environmental processes that are able to quench galaxies have been proposed. These include ram pressure stripping \citep[e.g.][]{Gunn72}, strangulation \citep[e.g.][]{Larson80,Balogh00}, galaxy-galaxy interactions and harassment \citep[e.g.][]{Mihos96,Moore98}, tidal interaction between the potential well of the environment and the galaxy \citep[e.g.][]{Merritt84,Fujita98}, or halo quenching \citep[e.g.][]{Birnboim03,Dekel06}; see \cite{Boselli06} for a review. The strength and quenching timescale of each physical process varies depending on many parameters such as the properties of the quenching environment (e.g. density, temperature, velocity dispersion, dynamical state) and those of the galaxy being quenched (e.g. mass, satellite vs. central). However, the ultimate consequence of each effect is to quickly or gradually remove the gas content/gas reservoir of galaxies or heat up the cold gas so that no further star formation activity is possible. 

There is substantial evidence for the truncation of the atomic and molecular gas content of cluster galaxies (especially in the outer parts of the disks), possibly due to ram pressure stripping (e.g. \citealp{Cayatte90,Boselli08,Fumagalli09}; see also \citealp{Boselli14} for a review). In principle, the dust distribution within galaxies should also depend on their host environment, most likely truncated outside-in as galaxies fall into the deeper potential wells of denser clusters. \cite{Cortese10} showed that atomic-gas-deficient spiral galaxies in the Virgo cluster also demonstrate signs of deficiency in their dust content. Similarly, \cite{Cortese12} found that the total dust content of atomic-gas-deficient spiral galaxies in Virgo is lower compared to normal field galaxies.
 
One might naively expect a continuous decline in the star formation and dust content of galaxies from the field to the dense cores of clusters. However, before galaxies undergo a full quenching process in dense regions, they may experience a temporary enhancement in star formation activity, which may complicate how observations are interpreted. For example, ram pressure stripping can initially compress the gas/dust which is favourable for star formation, and thus increasing the column density of the gas and dust \citep[e.g.][]{Gallazzi09,Bekki2009,Owers2012,Roediger2014}. Tidal galaxy-galaxy interactions can lead to the compression and inflow of the gas in the periphery of galaxies into the central part, feeding and rejuvenating the nuclear activity which results in a temporary enhancement in star formation activity \citep[e.g.][]{Mihos96,Kewley06,Ellison08}. Galaxy-galaxy encounters are more likely to happen when the interacting systems do not have extreme velocities (low-velocity-dispersion environment) and are closer to each other (denser regions). Intermediate-density environments such as galaxy groups, in-falling regions of clusters, cluster outskirts, merging clusters and galaxy filaments provide the ideal conditions for such interactions \citep[e.g.][]{Moss06,Perez09,Tonnesen12,Stroe2014,Stroe2015}. Therefore, one might expect a temporary enhancement in star formation activity and dusty star formation in intermediate-density environments before the galaxies quench. This has been found in several studies, referring to the intermediate-density environments as sites of enhanced star formation rate, star-forming fraction and obscured star formation activity \citep[e.g.][]{Smail99,Best04,Koyama08,Gallazzi09,Koyama10,Sobral11,Coppin12}. In practice, filaments may well be the actual dominant intermediate-density environment \citep[e.g.][]{Darvish14,Darvish15}.

The majority of studies rely on the far-infrared (FIR) or ultra-violet radiation (UV), radio/UV or FIR/H$\alpha$ ratio as an indicator of dust content of galaxies. \cite{Koyama11} performed an H$\alpha$ survey of the rich cluster Cl\,0939+4713 ($z=0.41$), and found a strong concentration of optically red star-forming galaxies in the group-scale environment around the cluster. They argue that the excess of the red star-forming galaxies suggests an enhancement of dust-obscured star formation in the group environment. By surveying the same cluster at $z=0.4$ with Spitzer (24\,$\mu$m), \cite{Koyama13} found that the 24\,$\mu$m to H$\alpha$ ratio increases from low to high density. This could be easily interpreted as an increase in dusty star-forming galaxies from low to high densities \citep[in line with e.g.][]{Rawle2012}. However, such high 24\,$\mu$m to H$\alpha$ ratios could also mean a higher fraction of AGN activity and/or declining star formation histories. This is because the FIR will measure SFRs over timescales of $\sim100$\,Myr, while H$\alpha$ is much more instantaneous ($\sim10$\,Myr). Measuring the Balmer decrement (H$\alpha$/H$\beta$) for star-forming galaxies only (after rejecting any potential AGN) would allow for significant progress.

In this paper, we present a spectroscopic follow-up of the H$\alpha$ candidate emitters in Cl\,0939+4713 found with Subaru and Suprime cam (S-Cam) and use data from \cite{Koyama13} to study their properties as a function of environment and internal properties. Section \ref{Sample} discusses the sample and its main properties. Section \ref{Observations} presents the observations with WHT/AF2, data reduction and emission line measurements. In Section \ref{Results} we show and discuss the results: the properties of individual galaxies, the stacked properties and the discussion of the results. Finally, Section \ref{Conclusion} presents the conclusions. We use AB magnitudes, a Chabrier \citep{Chabrier2003} initial mass function (IMF) and assume a cosmology with H$_{0}$=70kms$^{-1}$Mpc$^{-1}$, $\Omega_{M}$=0.3 and $\Omega_{\Lambda}$=0.7.

\section{Sample and properties} \label{Sample}

\subsection{The cluster} \label{Cluster}

The Cl\,0939+4713 cluster at $z=0.41$ (Abell 851) is one of the best-studied clusters at intermediate redshifts, and several studies have focused on the cluster central region(s) \citep[e.g.][]{Dressler_Gunn92, Dressler94,Stanford95,Smail99,Sato06,Dressler09,Oemler09}. Furthermore, wide-field ($\sim$30$'\times$30$'$) optical broad-band and narrow-band imaging surveys of this cluster have also been conducted \citep[e.g.][]{Kodama01,Koyama11} using S-Cam (\citealt{Miyazaki2002}) on the Subaru Telescope \citep{Iye2004}. Wide-field observations revealed 10\,Mpc-scale filamentary large-scale structures around the cluster based on the photometric redshift (photo-$z$) technique. Such structures around a massive cluster are similar to those found in other super-structures at $z\sim0.5-0.8$ \citep[e.g.][]{Sobral11,Darvish14,Darvish15}.

\subsection{Narrow-band survey and the sample of H$\alpha$ candidates} \label{NBsurvey}

\begin{figure}
\begin{tabular}{cccc}
\includegraphics[width=8.3cm]{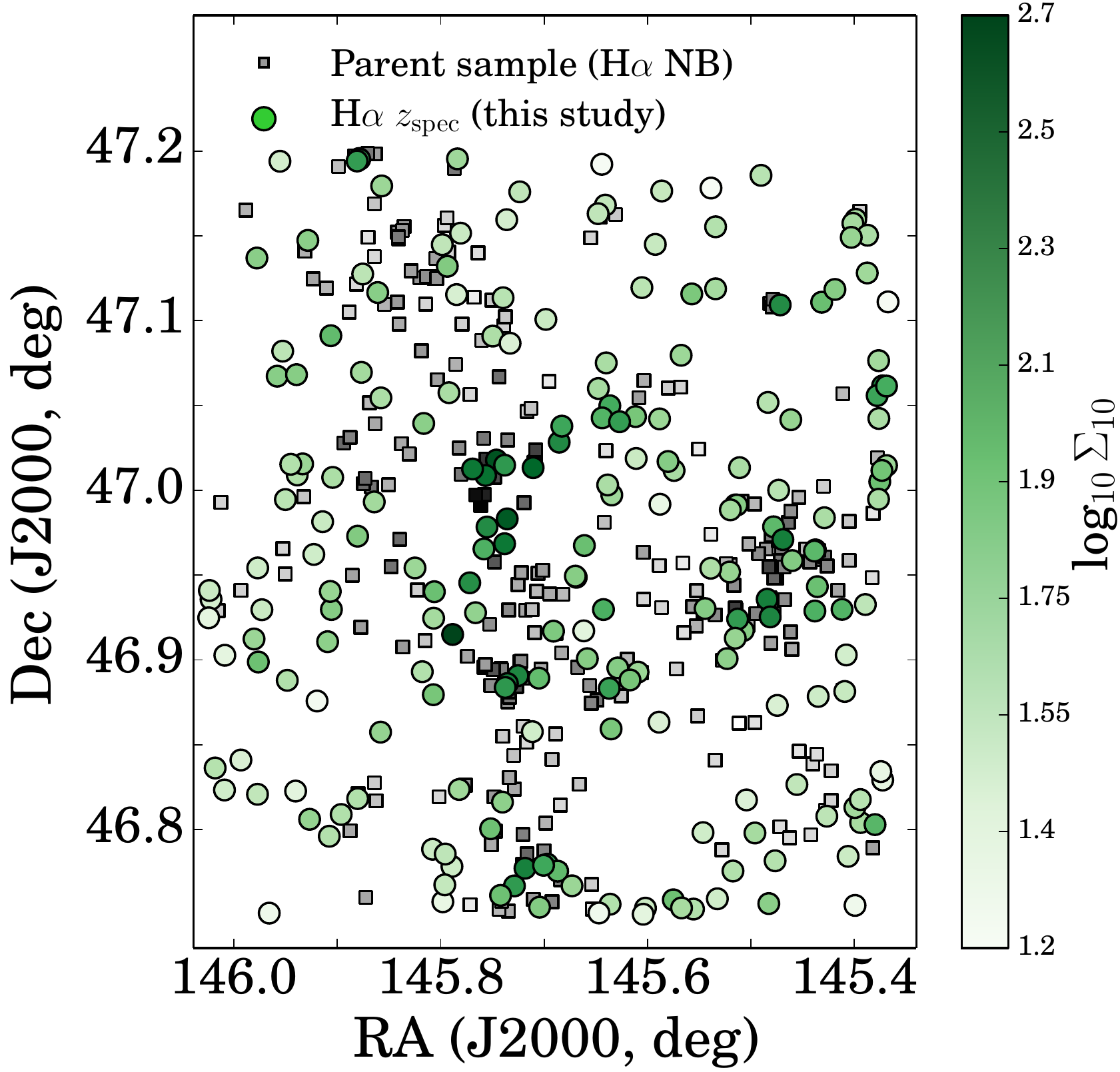}
\end{tabular} 
\caption{The on-sky distribution of the H$\alpha$ emitters from the parent sample and from our spectroscopic sample presented in this paper. We also show the local projected density of galaxies, from the lowest to the highest densities. Our spectroscopic sample probes the entire structure, including the highest densities, even though the fibre allocation for each single allocation makes it difficult to target the dense or crowded regions. For an alternative visualisation of the full environmental density range obtained with the full galaxy population, see \citealt{Koyama13}.}
\label{RADEC}
\end{figure}

\cite{Koyama11} conducted an H$\alpha$ narrow-band search around Abell 851. The authors used the NB921 filter ($\lambda_c=9196$\,\AA, $\Delta \lambda=132$\,\AA) on S-Cam and were able to identify $445$\,H$\alpha$ emitting galaxies in and around the cluster (see Figure \ref{RADEC}) down to a rest-frame equivalent width (EW), EW[H$\alpha$+[N{\sc ii}]] of $\sim20$\,\AA \ and a star formation rate (SFR) of $\sim0.3$\,M$_{\odot}$\,yr$^{-1}$. H$\alpha$ emitters are distinguished from other higher redshift emitters \citep[e.g.][]{Sobral13,Matthee15} using colour-colour selections \citep[see][for full details]{Koyama11}. In this paper, we follow-up 214 sources, a significant fraction ($\sim50$\%, see Figure \ref{RADEC}) of the 445 H$\alpha$ emitter candidates found by \cite{Koyama11}.

%
%
%
%
%

\subsection{Multiband photometry and stellar masses} \label{multiband_data}

We use multi-band catalogues derived by \cite{Koyama13} to obtain information on all the H$\alpha$ emitters: $z'$ band magnitude, $B-I$, $B-z'$, H$\alpha$ flux and luminosity from NB921, and EW(H$\alpha$+[N{\sc ii}]). By using $B-I$ and $z'$ band magnitude, \cite{Koyama13} derive stellar masses (which essentially uses $z'$ magnitudes and a colour). We follow the same methodology and derive masses by using: 

\begin{equation}
\log (M_*/10^{11}M_{\odot})_{[z=0.4]} = -0.4(z'-20.07) + \Delta\log M,
\end{equation}
where $\Delta\log M$:
\begin{equation}
\Delta\log M= 0.054 - 3.81\times \exp[-1.28\times (B-z')]-0.2.
\end{equation}

We note that the final term ($-0.2$) is added to the equation when compared to e.g. \cite{Koyama13} to convert from a Salpeter IMF to a Chabrier IMF, which we use. The typical (average) stellar mass of the full parent sample is 10$^{9.7\pm0.6}$\,M$_{\odot}$. For comparison, $M^*$ for star-forming galaxies at $z=0.4$ is $10^{10.9\pm0.1}$\,M$_{\odot}$ \citep[][]{Muzzin13,Sobral.14}. The reader is also referred to \cite{Koyama13} that shows that H$\alpha$ emitters in and around Abell 851 follow the same stellar mass-star formation rate relation as those in the general field \citep[][]{Sobral13,Sobral.14}.

We also use local environmental densities as derived in \cite{Koyama13}, along with cluster centric distances. We refer the reader to \cite{Koyama13} for more details on how the local environment densities were computed. Local environmental densities vary from log$_{10}$($\Sigma$)\,$\sim1.3$ to log$_{10}$($\Sigma$)\,$\sim2.7$. Briefly, local densities, log$_{10}$($\Sigma$), are calculated using all cluster member galaxies (photometric redshift selected and those which are H$\alpha$ selected) with the nearest-neighbour approach, calculated within a radius to the 10th-nearest neighbour from each source \citep[e.g.][]{Koyama11,Sobral11}. This method is in very good agreement with more robust density estimators such as adaptive smoothing and Voronoi tessellation \citep[][]{Darvish15b} and we use it throughout this work.

%
%
%
%
\begin{table*}
\caption{Observing log for the different nights and configurations used in this study. Observations were conducted with AutoFib2 (AF2) + Wide Field Fibre Optical Spectrograph (WYFFOS) on WHT using the R316R grism.}
\label{OBSERVATIONS}
\begin{tabular}{cccccccc}
\hline
Configuration 	&	 Exp. Time	&	Exp. Time & \# Targets	& Date  & Seeing & Sky & Moon \\
 (Pointing)   &  (Science)   	&	(Sky) 	& (Obs/Recov.)	 & (2015)  & ($''$)	& & \\
\hline 							  
P1 & 12.6\,ks  & 4.1\,ks  &	56 / 40  &   22 Jan    &	 1.2-1.5  & Clear & Dark  \\  
P2 & 13.5\,ks & 4.5\,ks &	55 / 39 &   23 Jan	 &	1-1.3  & Clear & Dark 	\\
P3 & 11.7\,ks & 4.1\,ks & 52 / 29 &    24 Jan	 &	1.5-2.0 & Clear & Dark 	\\
P4 & 12.6\,ks  & 4.1\,ks &	51 / 11  &    25 Jan	&	2-3 &  Clear & Dark  \\  
\hline 
\end{tabular}
\end{table*}

\section{Spectroscopic observations} \label{Observations}

\subsection{Observations: AF2 spectroscopy with WHT} \label{observations}

AutoFib2 (AF2; \citealt{Goodsell2003}) + Wide Field Fibre Optical Spectrograph (WYFFOS; \citealt{Dominguez14}) is a multi-object, wide-field, fibre spectrograph mounted at the Prime focus of the 4.2\,m William Herschel Telescope (WHT). The AF2+WYFFOS (AF2 for short, from now on) instrument on WHT is made of 150 science fibres, each with a diameter of 1.6$''$, which can be allocated to sources within a $\sim30\times30$\,arcmin$^2$ field of view, although with strong spatial constraints/limitations. We used the Red+4 detector and the R316R grism with a central wavelength of $\sim8000$\,\AA. The central wavelength varies slightly depending on the fibre and field location, but for a source at $z=0.4$ all our spectra cover the main emission lines we are interested in: [O{\sc ii}]\,3727\AA, H$\beta$, [O{\sc iii}]\,4959,\,5007\,\AA, H$\alpha$ and [N{\sc ii}]\,6584\AA. The R316R has a dispersion of 1.7\,\AA\,pix$^{-1}$ and a resolution of 8\,\AA, which corresponds to a resolution of just under 6\,\AA \ at $z=0.4$.

We followed up 214 candidate H$\alpha$ line emitters from \cite{Koyama11} using AF2 on the WHT in La Palma (program ID: W14BN020) on the second half of four nights during 2015 January 22--25 (see Table \ref{OBSERVATIONS}). Targets have estimated emission line fluxes at $\lambda\sim9196$\,\AA \ of $5\times10^{-17}$\,erg\,s$^{-1}$\,cm$^{-2}$ to $9\times10^{-15}$\,erg\,s$^{-1}$\,cm$^{-2}$ and estimated (rest-frame) EW (H$\alpha$+[N{\sc ii}]) in the range 20\,\AA \ to 700\,\AA. We targeted the cluster with 4 different configurations (P1--P4), all centred on the cluster, but with slightly different rotator angles and small offsets, in order to sample as best as possible the highest density regions, but also the regions at larger cluster centric distances. P1 targeted 56 H$\alpha$ candidates, P2 55, P3 52, and P4 targeted 51 H$\alpha$ candidates at $z\sim0.4$. In each configuration, fibres which were not possible to allocate to targets due to over-crowding were allocated to sky on the edges of the field of view.

The seeing was 1.0--1.5$''$ in the first two nights (for 2 of the configurations), being slightly worse ($\sim1.5-2''$) for the third night (third configuration), and becoming even worse (2--3$''$) for the final night (final configuration). Thus, the success rate for configurations P1 and P2 was higher than for P3 and much higher than P4 (see e.g. Table \ref{OBSERVATIONS}). The sky was clear for the entire observing run. We observed each configuration with individual exposure times of 900\,s or 450\,s, observing an offset sky position for the same exposure time for roughly every two science frames (AAB). In total, we obtained $\sim13$\,ks of science exposure time and $\sim4$\,ks of sky exposures per configuration. See Table \ref{OBSERVATIONS} for more details on the observations.

\subsection{AF2 data reduction} \label{reduction}

In order to reduce the data, we followed the same steps as described in \cite{Sobral2015}. We took biases and lamp flats both before and after the observations. Arcs were taken using neon, helium and mercury lamps with the fibres in the on-sky configuration; the same procedure was followed for flats. The traces of the fibres on the CCD were curved in the dispersion direction (y axis on the CCD) and the lamp flats were used to correct for this distortion. We fit each (flat) fibre shape with a Y pixel coordinate polynomial as a function of X coordinate. We correct all of the CCD pixels with the polynomial obtained for the closest fibre. We note that these corrections are applied to each configuration (P1--P4) separately. We also apply the corrections to the biases, flats, lamp arcs and the science data.

The final 2D bias subtracted and curvature corrected frames were then sky subtracted using the sky position exposure(s). In order to improve the sky subtraction we also used sky-dedicated fibres to scale the counts. Roughly 20 of such fibres were allocated to sky in each of the 4 pointings/configurations, allowing a robust scaling of the sky. We further obtained the best scaling factor by minimising the residuals after sky subtraction. After subtracting the sky, we extracted sources along the dispersion axis, summing up the counts. We extracted the signal on the flats, normalised the fibre response and flat fielded each spectrum in 1D. We obtained a first order wavelength calibration by using the arcs and obtain a final wavelength calibration per fibre by using the wealth of sky lines ($\sim100$ different lines, from $\lambda\sim5570$\AA \ to $\lambda\sim10000$\,\AA) on that particular fibre. We do this by extracting the co-added sky frames, and extract each spectrum in the same way as the science spectra. After matching all the sky lines for each fibre/spectrum individually, we obtain a wavelength calibration with an error (rms) of $\sim2$\,\AA \ (rest-frame $\sim1.4$\,\AA \ at $z\sim0.4$), of the order of the pixel scale of $\sim2$\,\AA\,pix$^{-1}$.

We further investigate the response curve to derive any corrections needed to robustly compare e.g. emission lines being detected at large separations in observed wavelength. We do this by both using stars with a flat spectrum, and also by constructing a flat spectrum by stacking the observed spectra without redshifting them and after masking all emission and absorption lines from the sources themselves. For a flat spectrum, we find a much higher response at the lowest wavelengths, making all sources look very blue. We derive corrections per pointing and apply them to each individual spectrum, by dividing the spectrum by our correction. We then check (for sources with sufficiently high signal to noise continuum) that our spectra yields colours consistent with the broad-band colours. We note that this correction also corrects for the telluric B-band absorption at $\sim6860-6890$\,\AA \ (although this would only affect $\sim10$\,\% of our sources, those with $z>0.41$).

%
%
%
%
\begin{table*}
\caption{A summary of the spectroscopic sample, for all the 214 sources followed-up. We find that 212 sources are consistent with being H$\alpha$ emitters at $z\sim0.4$, but we separate them in different groups based on the number of lines detected and on the significance of their detections. Flag 1: H$\alpha$ and at least 2 other emission lines; Flag 2: H$\alpha$ and at least another emission line detected at S/N$>5$; Flag 3: at least one emission line detected at S/N$>5$. Flag 4: a potential H$\alpha$ line but at too low S/N (S/N$<5$). Note that the vast majority of Flag 4 sources are from very low S/N observations due to poor seeing, or at the lowest H$\alpha$ fluxes.}
\label{flag_sample}
\begin{tabular}{cccccccc}
\hline
Configuration 	&	 Targets  & Non-H$\alpha$	&	Flag 1	& Flag 2  & Flag 3 & Flag 4 & Flags 1,2,3 \\
 (Pointing)   &  \#  &  \#  	&	H$\alpha$+2L 	& H$\alpha$+1L	 & 1L S/N$>5$ & H$\alpha$ low S/N	& In this paper  \\
\hline 							  
P1 & 56  & 1 & 29  & 2  &   9    &	 16  & 40   \\  
P2 & 55 & 1 & 26 & 2 &   11	 &	14  & 39 	\\
P3 & 52 & 0  & 17 & 2 &    10	 &	23 & 29  	\\
P4 & 51 & 0  & 0 & 1  &    10	&	40 &  11   \\ 
\hline 	
Full Sample & 212 & 2 & 72 & 7 &  40 & 93 &  119   \\   
\hline 
\end{tabular}
\end{table*}

\subsection{Redshifts} \label{Redshifts}

%
%

We were able to observe 214 out of our full sample of 445 potential H$\alpha$ emitters spread over an area which is very well matched to the AF2 field-of-view (see Figure \ref{RADEC}). The R316R grating enabled the de-blending of the H$\alpha$ and [N{\sc ii}] emission lines at the redshift of the cluster ($z=0.41$; separation of $\sim20$\AA \ with $\sim6$\,\AA \ resolution). Our spectra have a wide enough wavelength coverage to allow us to search for all main emission lines we need to unveil the nature of the emitters ([N{\sc ii}], H$\alpha$, [O{\sc iii}], H$\beta$ and [O{\sc ii}]).

%
%
%
\begin{figure}
\begin{tabular}{cccc}
\includegraphics[width=8cm]{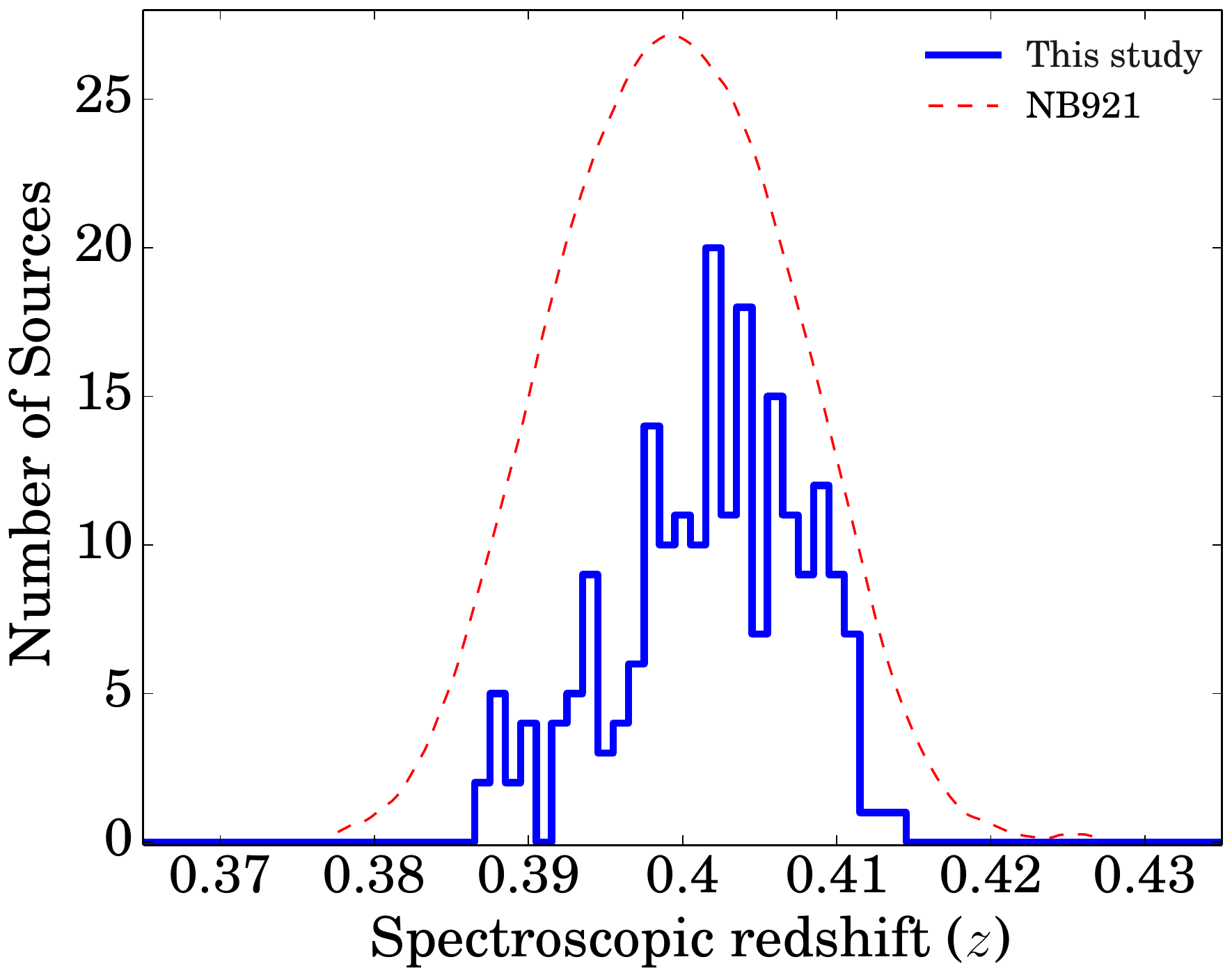}
\end{tabular} 
\caption{The distribution of spectroscopic redshifts obtained after following up the H$\alpha$ candidates in and around the cluster, compared to that expected based on the detection of the H$\alpha$ line with the NB921 filter on Subaru.}
\label{Histogram_zs}
\end{figure}

In order to obtain redshifts for the observed sources (Figure \ref{Histogram_zs}), we first search for an emission line at $\lambda\sim9196$\,\AA \ and, if that is found, assign a potential redshift assuming it is H$\alpha$. If at least two other major emission lines are found matching that redshift (e.g. [O{\sc ii}], H$\beta$, [O{\sc iii}] or [N{\sc ii}]), we measure those positions, determine the redshift for each one, combine the measurements, and we give the source a flag of 1, as the redshift is based on 3 lines or more. The redshifts of these sources are the most robust within our sample. If we find only another strong line, we flag the source with a 2. The redshift is still considered to be robust. These sources dominate the lower flux part of the sample. If we are not able to find any other emission line apart from that found at $\lambda\sim9196$\,\AA, we still assume it is H$\alpha$, but we flag these sources as 3. These sources are the ones with the lowest fluxes in our sample. Finally, if we fail to detect any line (this happens for cases where the estimated line flux is the weakest), we look for any other emission lines, and we follow the flag system according to the number of lines found. When no other emission lines are found and the only line found (at $\lambda\sim9196$\,\AA) is at S/N$<5$, the source is flagged as 4, and is not used at all in the analysis. There are 93 sources flagged as 4 (see Table \ref{flag_sample}), with the vast majority being sources targeted in configuration 4 (with the poorest seeing) and then configuration 3 (with the second poorest seeing). Sources from pointings 1 and 2 in this group are those with the lowest fluxes in the sample. Within our followed-up sources we find two (2) interlopers: a star and a lower redshift emitter ($z=0.13301$), with the remaining sources being very likely H$\alpha$ emitters with redshifts ranging from $z=0.3883$ and $z=0.4125$. Our final sample is made of 119 sources with flags 1, 2, 3, and with H$\alpha$ fluxes $>1.25\times10^{-16}$\,erg\,s$^{-1}$\,cm$^{-2}$ (see Table \ref{flag_sample}).

\begin{figure}
\includegraphics[width=8.3cm]{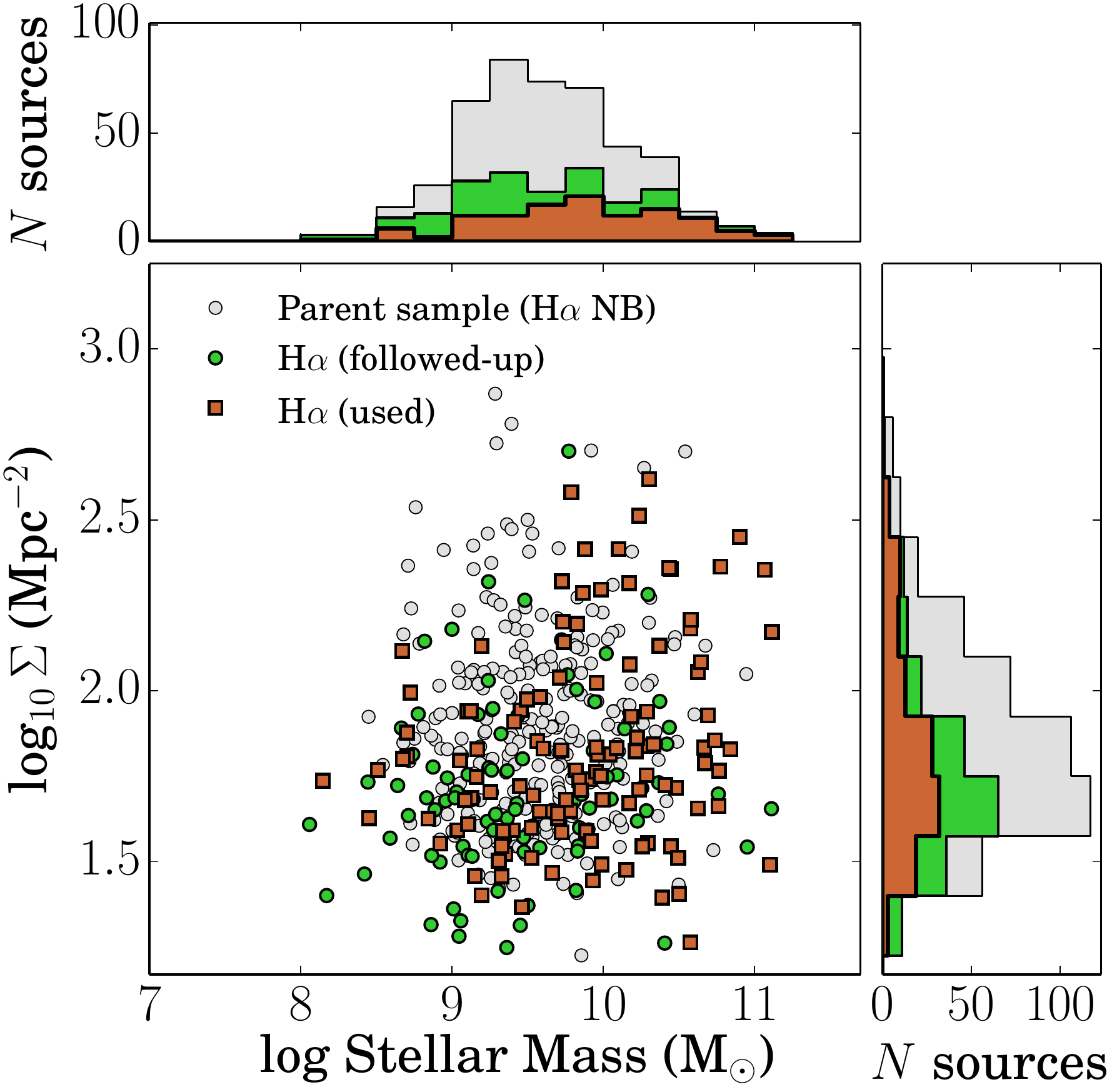}
\caption{The dependence of environmental density on stellar mass for our H$\alpha$ emitters. We show the full parent main sample of 445 H$\alpha$ emitters, and also the sample of H$\alpha$ emitters that we use in this paper, compared to all sources that were followed-up, including those that were followed-up in poor seeing conditions. We find a weak correlation between environmental density and stellar mass (at $\sim1-2$\,$\sigma$ level), in both the parent and our spectroscopic sample, with slopes of $\sim0.06$.}
\label{Density-mass}
\end{figure}

We present the redshift distribution of those $z\sim0.4$ sources in Figure \ref{Histogram_zs}. We also show the on-sky location of our spectroscopic sample in Figure \ref{RADEC}, which shows that, particularly due to the different configuration set-ups, we are able to probe the full range of environmental densities (see Figure \ref{Density-mass} and also \citealt{Koyama11}).

%
%
%
%

\subsection{Completeness and sample properties} \label{comple_sample}

\begin{figure}
\begin{tabular}{cccc}
\includegraphics[width=8.3cm]{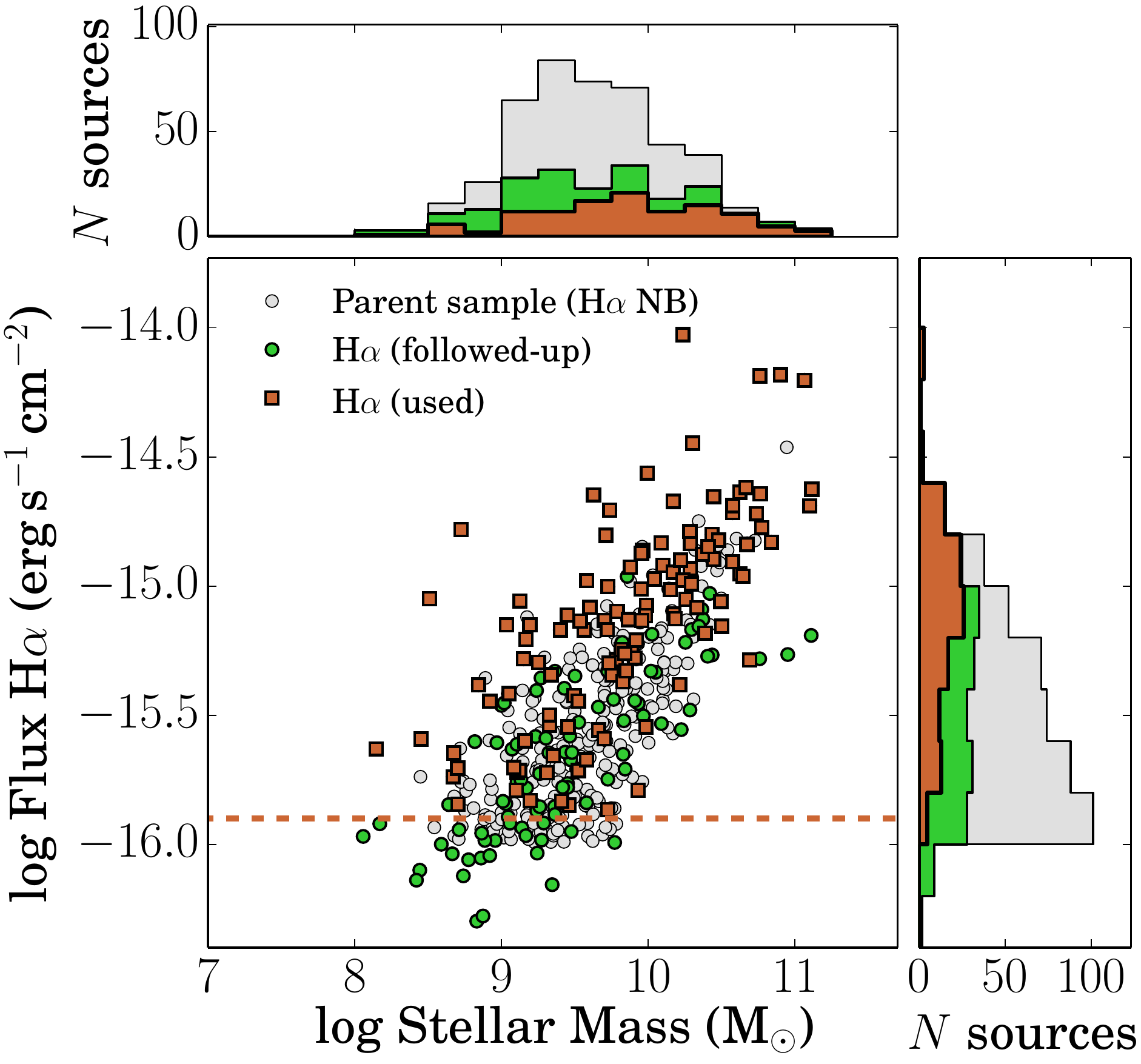}
\end{tabular} 
\caption{The dependence of H$\alpha$ flux (observed, from narrow-band, without any extinction correction) on stellar mass. We show the full parent main sample of 445 H$\alpha$ emitters, and also the sample of H$\alpha$ emitters that we use in this paper, compared to the sample that was followed-up in the 4 pointings/configurations. We find that H$\alpha$ flux correlates well with stellar mass, for both the parent and for our spectroscopic sample. We also find that our spectroscopic sample covers the full stellar mass range of the parent sample, and also the full range in fluxes of the parent sample. Due to our weighting in allocating the fibres, the sources with the highest fluxes were prioritised, leading to a much flatter H$\alpha$ flux distribution of the sample: this allows us to have very good statistics as a function of both H$\alpha$ flux and stellar mass for both luminous and less luminous sources, instead of being fully dominated by the most numerous faint sources in the parent sample. We note that we also introduce a flux cut where our sample would not be representative: this is indicated with the dashed line.}
\label{MS_comparison}
\end{figure}

Figure \ref{MS_comparison} shows the relation between H$\alpha$+[N{\sc ii}] flux (based on narrow-band photometry, so we can compare it with the parent NB sample) and stellar mass, for both the parent sample, and for our spectroscopic sample. The comparison with the parent sample shows that our sample recovers the full range of masses and H$\alpha$ fluxes, even down to the lowest masses and also down to the lowest fluxes. Therefore, we apply no corrections, as the spectroscopic sample is consistent with being drawn from the parent sample. We note that because of the weighting we applied when assigning the fibres, we were able to obtain a flux distribution which is much flatter than the parent sample: this allows us to have good statistics at all masses and at all H$\alpha$ fluxes/luminosities, instead of being fully dominated by e.g. the much more numerous faint H$\alpha$ emitters.

%
%
%
%

We also show the distribution of stellar masses for the samples of H$\alpha$ emitters in Figures \ref{Density-mass} and \ref{MS_comparison}. As a whole, the sample of H$\alpha$ emitters at $z=0.41$ has a similar stellar mass distribution to samples of field H$\alpha$ emitters at similar redshifts \citep[see][]{Sobral.14}, but with cluster H$\alpha$ emitters having higher stellar masses than H$\alpha$ emitters outside the cluster \citep[see][]{Koyama13}.

\subsection{SFGs vs AGN in A851}  \label{AGN_SFG}

\begin{figure}
  \centering
 \includegraphics[width=8.5cm]{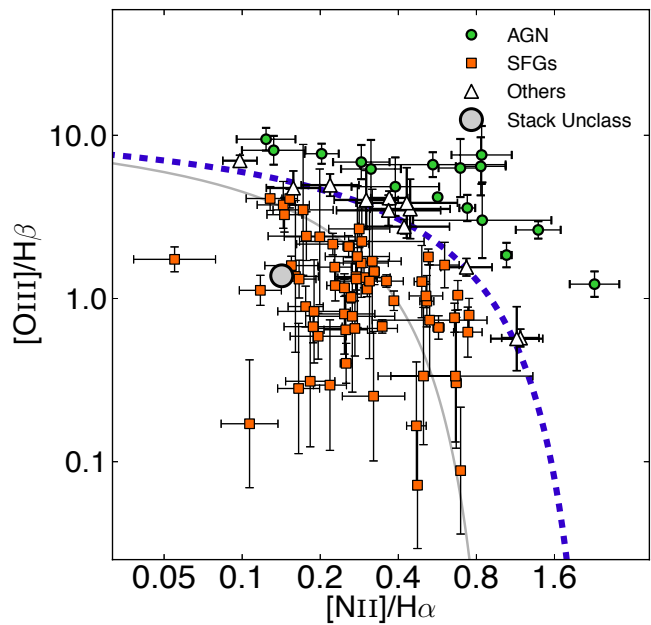}
\caption{Emission line ratio diagnostics \citep[][]{BPT} separate star-forming dominated from AGN dominated H$\alpha$ emitters. We show the location of pure, ``typical'' star-forming galaxies (grey solid line), and the separation line between maximal starbursts and AGN (blue dashed line) from \citet{Kewley01}.}
\label{AGNvsSF}
\end{figure}

Emission line fluxes are measured by fitting Gaussian profiles, and estimating the continuum directly red-ward and blue-ward of the lines (masking any other features or nearby lines). We also measure the rms directly blue- and red-ward of the lines, and assign that as the noise (1\,$\sigma$). In order to differentiate between star-forming and AGN, the [O{\sc iii}]\,5007/H$\beta$ and [N{\sc ii}]\,6583\AA/H$\alpha$ line ratios are used (see Figure \ref{AGNvsSF}); these have been widely used to separate AGN from star-forming galaxies \citep[e.g.][]{BPT,Rola,Kewley01,Kewley13}. These line ratios are measured for emission lines sufficiently close to each other that dust extinction has little effect.

We note that particularly H$\beta$ can be affected by significant underlying stellar absorption and that this can be a very important effect for galaxies with relatively low EW emission lines, more characteristic of e.g. mass-selected samples, which pick up galaxies with relatively low star formation activity. By definition our sources are all high EW sources and highly star-forming. The effect of the absorption is expected to be small given the large EW. Nevertheless, we further investigate the potential effect and assess whether a correction should be applied. We use the results from \cite{Darvish15} which explore a redshift regime, sample and selection which are very similar to those presented in our study. The advantage is that \cite{Darvish15} spectra are much higher resolution and significantly deeper (as they are obtained using DEIMOS on Keck). \cite{Darvish15} showed that H$\beta$ Balmer stellar absorption only accounts for 5-10\% correction to the H$\beta$ flux. Even the maximum 10\% correction results in only a $<0.05$\,dex difference in the line ratios, smaller than our observational uncertainties. Other studies \citep[e.g.][]{Zahid2011,Reddy2015} also find comparable corrections for similar samples of emission-line galaxies. Given the small corrections, we decide not to correct for Balmer absorption. We note, nonetheless, that our Balmer decrements as a function of e.g. stellar mass agree very well with \cite{Garn2010}, another indication that the correction, if any, must be relatively small. However, on a source by source basis, and particularly for relatively noisy spectra, H$\beta$ may be underestimated, leading to an overestimation of [O{\sc iii}]/H$\beta$. This may be happening to some of the sources above the maximal starburst line and with [N{\sc ii}]/H$\alpha<0.4$.

Figure \ref{AGNvsSF} shows data-points for the line ratios. We only show sources for which H$\alpha$ is detected at S/N$>5$ and for which we can measure emission line ratios with S/N$>2$, i.e. for 89 out of 119 sources in our sample. We investigate the remaining 30 sources in our sample separately by stacking them.

\begin{figure}
\includegraphics[width=8.44cm]{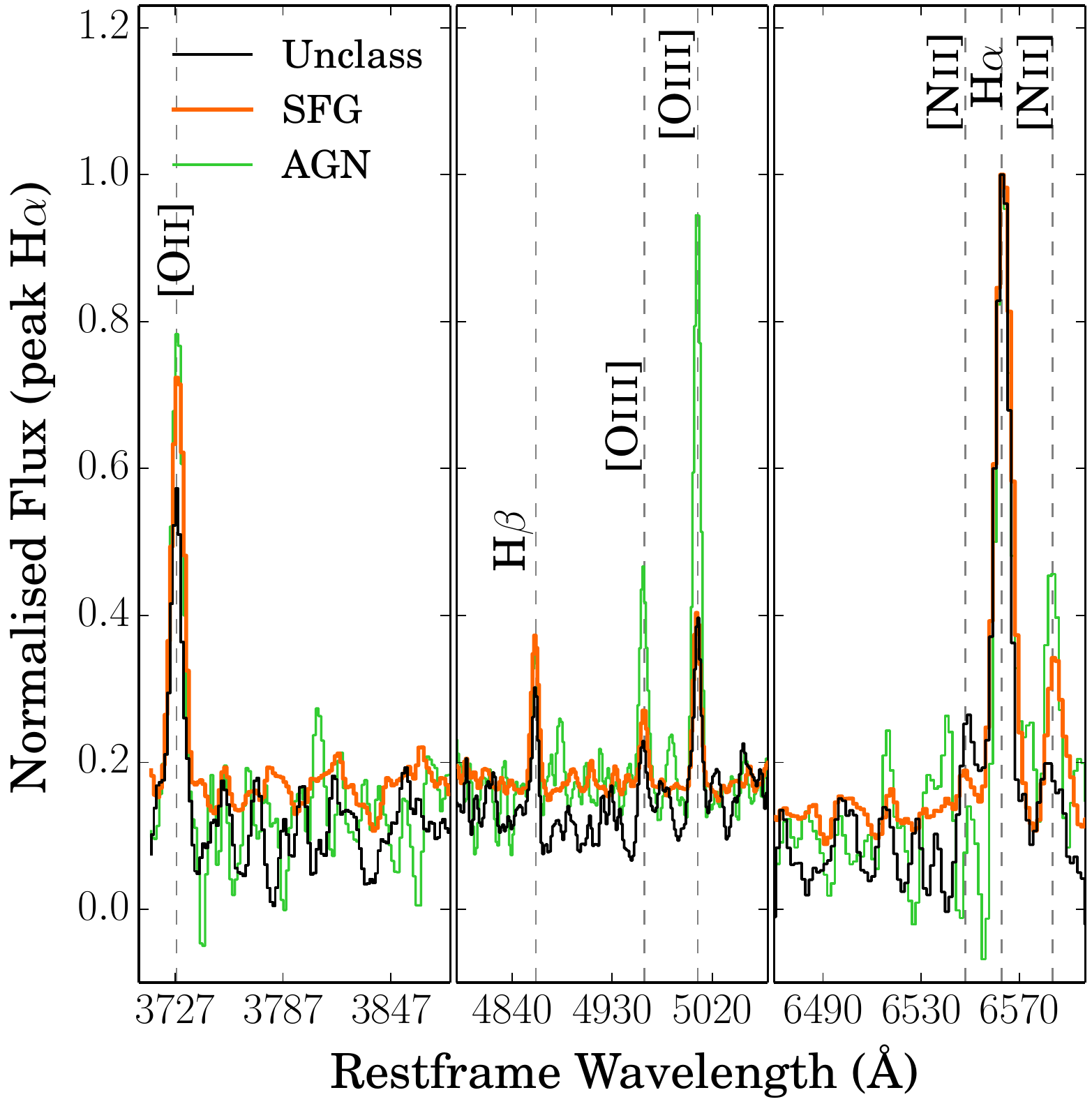}
\caption{Stacked spectra of sources dominated by different ionising sources: active galactic nuclei (AGN), star-forming galaxies (SFG) and those for which it is not possible to obtain individual line ratios (low S/N) and thus we label as `Unclass'. The stacks clearly reveal the strong differences between star-forming galaxies and AGN, particularly regarding the [O{\sc iii}] and [N{\sc ii}] emission lines (in respect to H$\beta$ and/or H$\alpha$): AGN have very strong [O{\sc iii}] emission, comparable to H$\alpha$, while star-forming galaxies have [O{\sc iii}] emission lines which are of the same order as H$\beta$. The stack of `Unclass' sources reveals line ratios which are in excellent agreement with being dominated by star-forming galaxies, but being a likely mix of lower mass (lower continuum), lower metallicity, and slightly dustier star-forming galaxies (higher H$\alpha$/H$\beta$). For the remaining of the analysis, we consider these galaxies as star-forming.}
\label{STACK_SPECTRA}
\end{figure}

Figure \ref{AGNvsSF} shows curves from \cite{Kewley01} and \cite{Kewley13} encompassing star-forming galaxies (grey solid line), and encompassing up to maximal starbursts (blue dashed line). We use the blue dashed line to separate between what we now refer as AGN and star-forming galaxies. We find that out of the galaxies that we can classify (89 sources), 17\% of our H$\alpha$ emitters are consistent with having AGN activity (15 sources, 13\% of full sample), while 70\% (62 sources) are star-forming galaxies (see Table \ref{Summary_final}). From the remaining, 12 sources are too close to the boundary between AGN and star-forming (see Figure \ref{AGNvsSF}), so we label them as `others' and we do not use them for the rest of the analysis. Figure \ref{STACK_SPECTRA} shows how the median stacked spectra of our AGN compares with a similar stack for our star-forming galaxies, clearly revealing the strong median differences of line ratios and line fluxes. Our stacks are all normalised to the peak of the H$\alpha$ emission.

In order to further investigate the nature of the non-classifiable sources (30 sources), we stack them and measure their median properties. We stack spectra by normalising them to the peak of the H$\alpha$ emission. We show the results in Figure \ref{STACK_SPECTRA} and compare them with equally stacked spectra of our clear star-forming galaxies and our AGN H$\alpha$ emitters. We find line ratios which are completely consistent with a likely mix of relatively metal-poor, low mass, and moderately dust-extinguished star-forming galaxies, with lower fluxes. Our lower S/N sources show emission line properties and ratios which clearly indicate a star-forming nature, at least for the bulk of such sample. From now on, we will use our sample of clear star-forming galaxies, and our low S/N sample that we have found to be consistent with star-formation. Thus, our final star-forming sample is made of 92 likely star-forming galaxies, after rejecting AGN and those sources which are too close to the AGN selection region (see Table \ref{Summary_final}).

%
%
%
%

%
%
%
%
%

\section{Results and Discussion} \label{Results}

\subsection{AGN fraction as a function of environment}  \label{AGN_SFG}

%
%
%
%
\begin{table}
\centering
\caption{A summary of the final sample and the likely ionising sources: SFGs or AGN. For sources too close to the boundary between SFGs and AGN, we label them ``Others", and we exclude them from the analysis of SFGs. There are sources (Unclass) for which it is not possible to obtain reliable measurements of the required emission lines to classify them. We instead stack them, finding that they are fully consistent with being lower mass, lower flux SFGs.}
\label{Summary_final}
\begin{tabular}{cc}
\hline
Classification 	&	 Number  \\
\hline
Full sample & 119 (100\%)  \\ 
\hline					  
SFGs & 62 (52\%)  \\  
AGN & 15 (13\%)  	\\
Others & 12 (10\%)  	\\
Unclass & 30 (25\%)   \\ 
\hline 	
SFGs used & 92 (77\%)  \\  
\hline 
\end{tabular}
\end{table}

\begin{figure}
\centering
\includegraphics[width=8cm]{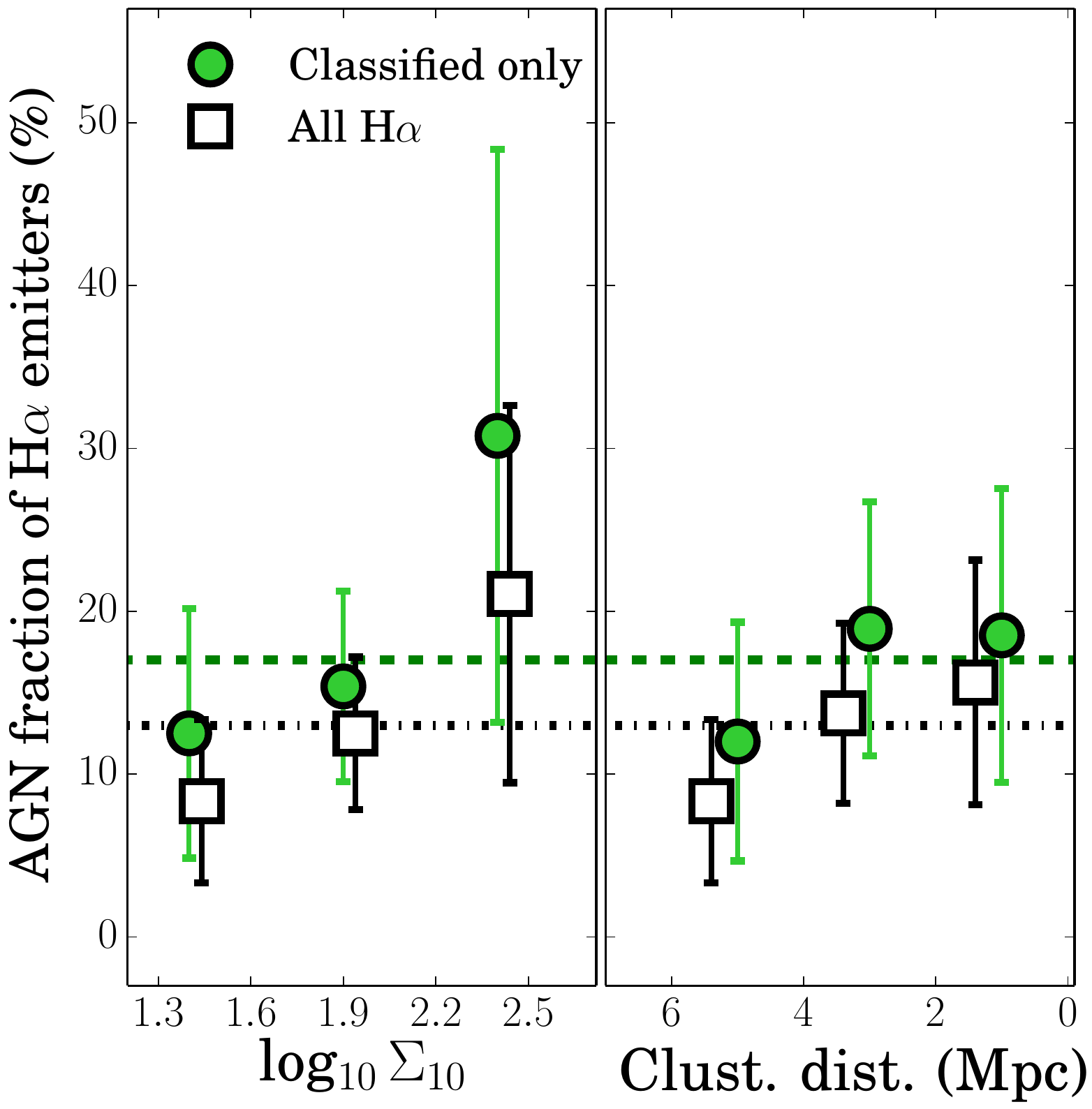}
\caption{{\it Left}: AGN fraction as a function of local density. We show AGN fractions when using the sample of classified sources only (89 sources) and when using the full sample of 119 sources. We find a very weak increase of the AGN fraction with increasing local density, at $<1$\,$\sigma$. {\it Right}: AGN fraction as a function of projected distance from the cluster centre, showing a relatively constant AGN fraction, with a very weak decline at the highest distances from the cluster ($<1$\,$\sigma$). Lines show the average AGN fraction for either classified or for all H$\alpha$ emitters.}
\label{AGN_FRACTIONS}
\end{figure}

In order to investigate the prevalence of AGN among H$\alpha$ emitters in and around the cluster, we evaluate the AGN fraction for different environmental densities and also as a function of cluster-centric distance. The results are presented in Figure \ref{AGN_FRACTIONS}. We find that the AGN fraction shows no significant correlation with increasing projected local number density nor with increasing projected cluster-centric distance (Figure \ref{AGN_FRACTIONS}). There is perhaps a very weak trend of increasing AGN fraction with environmental density, but for all cases at the level of $<1$\,$\sigma$, and thus not statistically significant. This is qualitatively and quantitatively consistent with e.g. \cite{Miller2003}, who found an AGN fraction of $\sim$20\,\% in the local universe, independent of environment.

Our results therefore show that there is no particular prevalence of AGN at high densities and/or close to the cluster that could explain the high 24\,$\mu$m/H$\alpha$ as a simple consequence of higher AGN fraction at higher densities. Thus, two main explanations for the elevated 24\,$\mu$m/H$\alpha$ as a function of environmental density for H$\alpha$ emitters still remain: dust extinction correlating with environmental density, or witnessing declining star formation histories, and thus, effectively, `slow' environmental quenching. In the following sections we evaluate the properties of the star-forming population in order to identify which explanation is the most likely.

%
%
%
%
%

\subsection{Properties of SFGs in and around A851 from galaxy by galaxy measurements}  \label{Gen_properties}

\subsubsection{Balmer decrement: Dust extinction} \label{balmer_indi}

The emission line ratio between H$\beta$ and H$\alpha$ is one of the most robust means of estimating dust extinction (A$_{\rm H\alpha}$) within star-forming galaxies. The H$\alpha$/H$\beta$ line ratio (Balmer decrement) is widely used as an extinction estimator, particularly up to $z\sim0.4$, as it is relatively easy to obtain both emission lines. H$\alpha$/H$\beta$ line fluxes are measured and used to estimate the extinction at H$\alpha$, A$_{\rm H\alpha}$, by using:

\begin{equation}
A_{\rm H\alpha} = \frac{-2.5 k_{\rm H\alpha}}{k_{\rm H\beta}-k_{\rm H\alpha}}~{\rm log}_{10}\left(\frac{2.86}{{\rm H\alpha}/{\rm H\beta}}\right),
\end{equation}
where 2.86 is the assumed intrinsic H$\alpha$/H$\beta$ line flux ratio, appropriate for Case B recombination, temperature of $T = 10^{4}$~K and an electron density of $n_{\rm e} = 10^{2}$~cm$^{-3}$ \citep{Brocklehurst71}. The \citet{Calzetti00} dust attenuation law is used to calculate the values of $k_{\lambda} \equiv A_{\lambda} / E(B-V)$ at the wavelengths of the H$\alpha$ and H$\beta$ emission lines, resulting in:

\begin{equation}
A_{\rm H\alpha} = 6.531\log_{10} \rm H\alpha/H\beta-2.981.
\end{equation}

Individual measurements for high S/N H$\alpha$ detections provide an average of $A_{H\alpha}=0.9\pm0.9$. We also find a relation in very good agreement with \cite{Garn2010}, with dust extinction increasing with stellar mass. For individual detections, we do not find a significant trend of dust extinction with environment. This could lead to the potential conclusion that the elevated 24\,$\mu$m/H$\alpha$ ratio may not be due to increasing dust extinction. However, using sources for which we can measure the Balmer decrement individually biases our results towards the less dust extinguished sources at all environmental densities. This can only be solved by means of stacking -- see \S\ref{balmer_indi}.

\subsubsection{Metallicities}  \label{Metals}

In order to investigate the metallicities of the ionised gas within the H$\alpha$ emitters in and around A851, we use both [N{\sc ii}]/H$\alpha$, but also O3N2. We only compute individual (source by source) metallicities for our H$\alpha$ emitters in and around the cluster that have been clearly identified as star-forming galaxies. We note that in \S \ref{stacks}, when stacking, we remove the AGN and galaxies unclassified due to being to close to the AGN selection function, but use the lower S/N galaxies, consistent with being dominated by star formation, so that we do not bias our results.

We start by using the [N{\sc ii}]/H$\alpha$ emission line ratio. For our full sample (median stack), we find [N{\sc ii}]\,/\,H$\alpha$\,=\,$0.33\pm0.19$ (see e.g. Figure \ref{STACK_SPECTRA}). The [N{\sc ii}]/H$\alpha$ line ratio can be used to obtain the metallicity of our star-forming galaxies (oxygen abundance), [12\,+\,log(O/H)], by using the conversion of \cite{Pettini04}:

\begin{equation}
\rm 12 + \log10(O/H)=8.9+0.57\times\log_{10}(N2H\alpha), 
\end{equation}
where N2H$\alpha$ is the line flux ratio [N{\sc ii}]6584/H$\alpha$. The galaxies in our full sample ([N{\sc ii}]/H$\alpha$ = $0.33\pm0.19$) have a median metallicity 12\,+\,log(O/H)\,$=8.59\pm0.19$, which is consistent with solar (8.66$\pm$0.05), but we note that we are sampling galaxies with a large range of masses, and that we find a strong mass-metallicity relation, and thus we need to take that into account when properly comparing the samples.

We also use the O3N2 indicator \citep[e.g.][]{Alloin79,Pettini04} as a tracer of metallicity (the gas-phase abundance of oxygen relative to hydrogen), computed by using:
\begin{equation}
\rm 12+\log10(O/H)=8.73-0.32\times\log_{10}(O3H\beta/N2H\alpha), 
\end{equation}
where O3H$\beta$ is the line flux ratio [O{\sc iii}]5007/H$\beta$ and N2H$\alpha$ is the line flux ratio [N{\sc ii}]6584/H$\alpha$. This indicator has the main advantages of i) using emission lines which have very similar wavelengths, thus being essentially independent of dust attenuation and ii) having a unique metallicity for each line flux ratio. 

By using O3N2, we find an average metallicity of 12\,+\,log(O/H)\,$=8.56\pm0.15$, thus consistent with solar and completely consistent with the [N{\sc ii}]/H$\alpha$ metallicity for the entire sample of star-forming galaxies. For the remaining of the analysis in the paper, we will use O3N2 metallicities.

\subsubsection{Mass-metallicity relation}  \label{mass_metal}

\begin{figure}
\centering
\includegraphics[width=8cm]{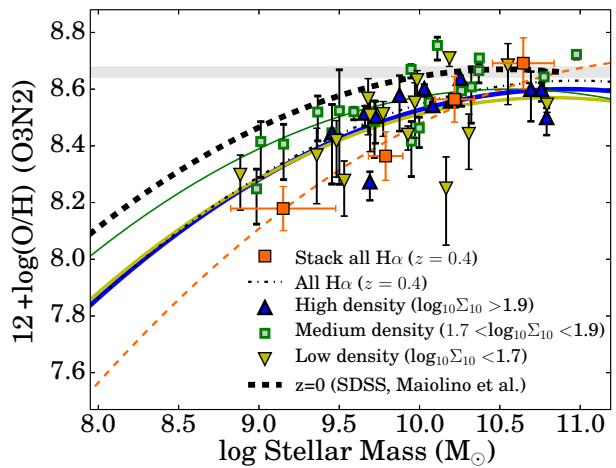} 
\caption{The mass metallicity relation for our full sample, for star-forming galaxies where we can measure all emission lines and obtain an O3N2-based metallicity. We also show the results when stacking, which include sources for which it  is not possible to measure all the lines individually. Our H$\alpha$ emitters show a clear mass-metallicity relation, with evolution when compared with the local Universe. We also split the sample in different densities and fit mass-metallicity relations. We find that while they are within $\approx1$\,$\sigma$, there is a slight trend of higher metallicity for intermediate-density environments, but the trend is very mild. Overall, our $z=0.4$ H$\alpha$ emitters show a mass metallicity which is lower by $\approx0.2$\,dex from the local Universe.}
\label{Mass_Metallicity}
\end{figure}

We find a mass metallicity relation for both [N{\sc ii}]/H$\alpha$ and O3N2 calibrations. The O3N2 metallicities provide a slightly steeper relation with mass, with a linear fit 12\,+\,log(O/H,O3N2)$=0.120\times\log_{10}M+7.357$, while N2Ha provides 12\,+\,log(O/H,N2Ha)$=0.074\times\log_{10}M+7.861$. Our results are shown in Figure \ref{Mass_Metallicity}.

In Figure \ref{Mass_Metallicity}, apart from showing how metallicity correlates with stellar mass for individual star-forming galaxies in and around the cluster, we also split our star-forming galaxies according to their local densities. We thus divide the sample into three sub-samples: low density, medium density and high density. We fit mass-metallicity relations to each of the sub-samples. While we find that star-forming galaxies residing in the highest local densities have slightly higher stellar masses \citep[consistent with e.g.][]{Koyama13,Sobral.13b}, we find that the mass-metallicity relation is the same, within the uncertainties, for all environments. We thus conclude that, for our sample of high S/N star-forming galaxies, the mass-metallicity is independent of the local environment. The results are consistent with \cite{Sobral.13b} at $z\sim0.8$.

\subsubsection{Ionisation parameter} \label{ion_param}

We compute the ionisation parameter ($q_{\rm ion}$) for our sample of H$\alpha$ emitters as a whole. In order to do this, we use the ratio [O{\sc iii}]/[O{\sc ii}]. We use equation 5 from \cite{NakajimaOuchi2014} which provides $q_{\rm ion}$ as a function of [O{\sc iii}]/[O{\sc ii}] and also takes into account the metallicity, 12\,+\,log(O/H,R23)\footnote{We correct our O3N2 metallicities to R23 by following \cite{KewleyEllison2008}.}:
\begin{eqnarray}
   && \log(q_{\rm ion})=\{ 32.81-1.153y^2 \nonumber\\
   && \ \ \ +[12+\log({\rm O/H})](-3.396-0.025y+0.1444y^2) \} \nonumber\\
   && \ \ \ \times \{ 4.603-0.3119y-0.163y^2 \nonumber\\
   && \ \ \ +[12+\log({\rm O/H})](-0.48+0.0271y+0.02037y^2) \}^{-1}, \nonumber\\
   &&
\end{eqnarray}
%
where $y=\log_{10}$([O{\sc iii}]/[O{\sc ii}]), and in this case [O{\sc iii}] is the sum of the flux of [O{\sc iii}]\,4959 and [O{\sc iii}]\,5007. We find a median of log($q_{\rm ion}$) = $7.3\pm0.3$, consistent with e.g. SDSS \citep[7.18-7.51][]{NakajimaOuchi2014}. We evaluate the ionisation parameter as a function of environmental density and stellar mass in \S\ref{stacks}.

%

%
%
%
%
%

%
%
%
\begin{table*}
\begin{center}
{\scriptsize
\caption{Results from our stacking analysis. We show the results for the stack of our full sample, and we further split our sample in respect to the likely ionising source. For AGN sources we do not calculate metallicities, dust extinction nor star formation rates, due to H$\alpha$ being potentially contaminated by AGN activity. We split the sample in respect to stellar mass and environmental density.} \label{table:RESULTS}
\begin{tabular}{cccccccccc}
\hline
\noalign{\smallskip}
SAMPLE   & \# & [N{\sc ii}]/H$\alpha$ & [O{\sc ii}]/H$\alpha$ & [O{\sc iii}]/H$\alpha$ & [O{\sc iii}]/[O{\sc ii}] & Mass & SFR     & 12\,+\,log(O/H) & A$_{\rm H\alpha}$   \\
  &  &  &  &  &  & log$_{10}$\,M$_{\odot}$ &  M$_{\odot}$\,yr$^{-1}$   & (O3N2)  & mag \\ 
\hline
Full Sample & 119 & $0.19\pm0.03$ & $0.73\pm0.02$ & $0.34\pm0.01$ & $0.47\pm0.03$ & $9.9\pm0.6$ & $0.7\pm0.4$ & $8.45\pm0.03$ & $1.06\pm0.08$  \\
  \hline
 AGN only & 15 & $0.57\pm0.11$ & $1.44\pm0.14$ & $1.25\pm0.12$ & $0.87\pm0.03$ & $9.8\pm0.7$ & --- & --- & --- \\
AGN and others & 27 & $0.37\pm0.06$ & $1.0\pm0.06$ & $1.07\pm0.06$ & $1.07\pm0.02$ & $9.7\pm0.7$ & --- & --- & --- \\
All SFGs & 92 & $0.17\pm0.02$ & $0.68\pm0.02$ & $0.24\pm0.01$ & $0.36\pm0.05$ & $9.9\pm0.6$ & $0.7\pm0.4$ & $8.48\pm0.03$ & $1.04\pm0.08$  \\
  \hline
$8.6<\log_{10}$\,M\,$<9.6$ & 26 & $0.03\pm0.03$ & $0.62\pm0.04$ & $0.48\pm0.01$ & $0.78\pm0.07$ & $9.3\pm0.3$ & $0.0\pm0.2$ & $8.18\pm0.18$ & $0.35\pm0.15$ \\
$9.6<\log_{10}$\,M\,$<10.0$ & 21 & $0.07\pm0.02$ & $0.98\pm0.03$ & $0.28\pm0.01$ & $0.28\pm0.18$ & $9.8\pm0.1$ & $0.5\pm0.3$ & $8.36\pm0.06$ & $0.7\pm0.13$ \\
$10.0<\log_{10}$\,M\,$<10.4$ & 24 & $0.26\pm0.04$ & $0.78\pm0.03$ & $0.2\pm0.01$ & $0.26\pm0.09$ & $10.2\pm0.1$ & $0.9\pm0.3$ & $8.56\pm0.03$ & $1.16\pm0.11$ \\
$10.4<\log_{10}$\,M\,$<11.0$ & 19 & $0.3\pm0.05$ & $0.37\pm0.02$ & $0.08\pm0.01$ & $0.22\pm0.59$ & $10.6\pm0.2$ & $1.2\pm0.2$ & $8.69\pm0.03$ & $1.53\pm0.18$ \\
  \hline
$1.2<\Sigma<1.5$ & 13 & $0.41\pm0.07$ & $0.91\pm0.07$ & $0.18\pm0.02$ & $0.2\pm0.63$ & $10.0\pm0.6$ & $0.4\pm0.4$ & $8.67\pm0.03$ & $0.57\pm0.31$ \\
$1.5<\Sigma<1.75$ & 27 & $0.01\pm0.02$ & $0.67\pm0.02$ & $0.34\pm0.01$ & $0.5\pm0.07$ & $9.7\pm0.6$ & $0.5\pm0.4$ & $8.08\pm0.43$ & $0.77\pm0.12$ \\
$1.75<\Sigma<2.1$ & 35 & $0.26\pm0.03$ & $0.55\pm0.02$ & $0.28\pm0.01$ & $0.5\pm0.07$ & $10.0\pm0.6$ & $0.7\pm0.3$ & $8.53\pm0.03$ & $0.96\pm0.11$ \\
$2.1<\Sigma<2.6$ & 16 & $0.21\pm0.04$ & $0.63\pm0.03$ & $0.21\pm0.01$ & $0.33\pm0.13$ & $10.1\pm0.5$ & $1.1\pm0.3$ & $8.51\pm0.04$ & $1.56\pm0.16$ \\
  \hline
\end{tabular}
}
\end{center}
\end{table*}

%
%
%
%
\begin{figure}
  \centering
 \includegraphics[width=8.2cm]{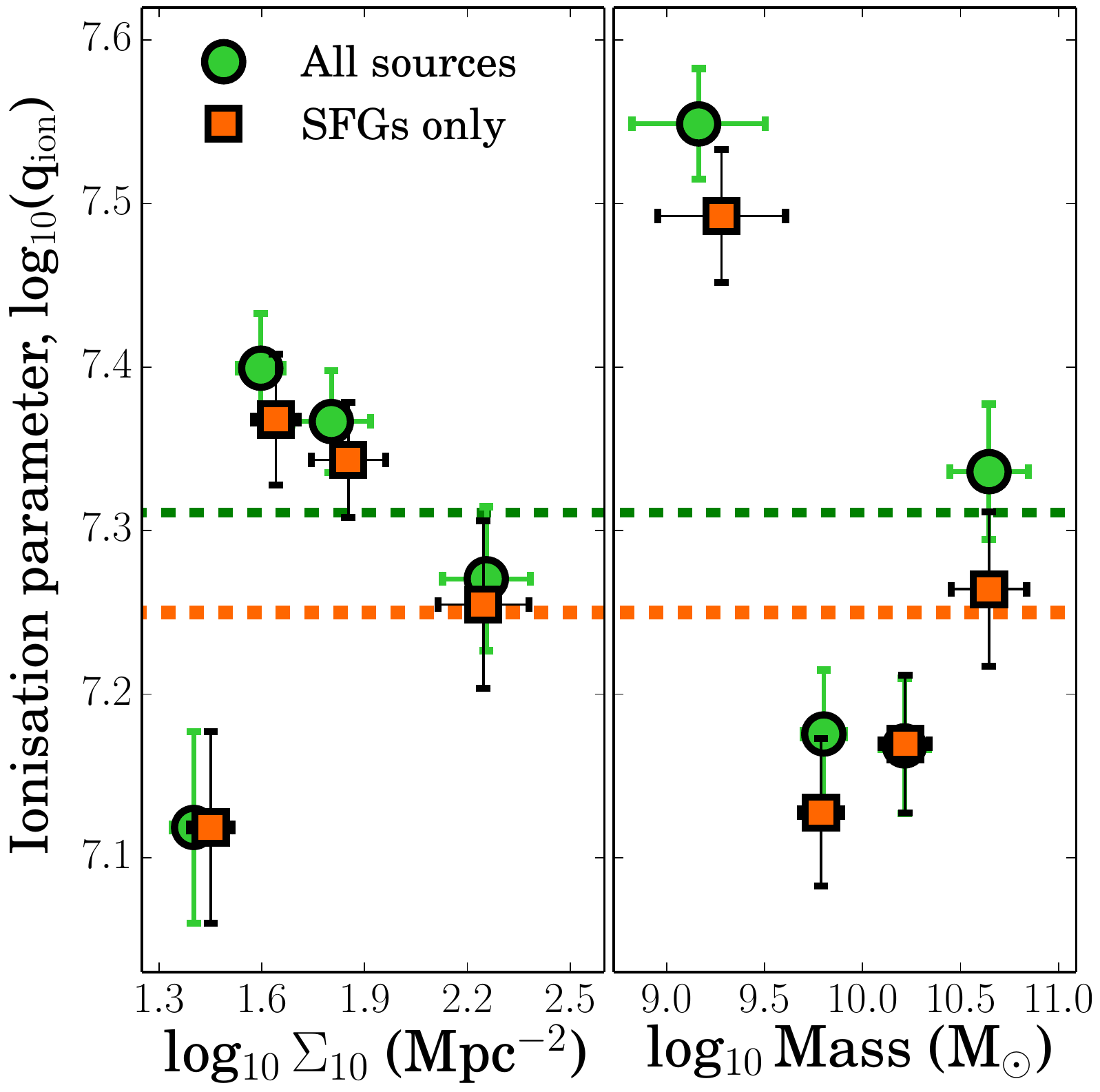}
\caption{Ionisation parameter from stacks as a function of environmental density ({\it left}) and stellar mass ({\it right}). Dashed lines indicate the stacked values for all sources and for star-forming galaxies only. We obtain stacks using our full sample (yielding higher ionisation parameters, due to the contribution of AGN) and using SFGs only. We find a potential rise and fall of the ionisation parameter with environment, but this is at least partially driven by the relation we find between stellar mass and ionisation parameter. The ionisation parameter has a much clearer and simple relation with stellar mass, being the highest for the lowest stellar masses, and declining with increasing stellar mass.}
\label{F_ion_vs_Dens_MASS}
\end{figure}

\subsection{Properties of SFGs in and around A851 from stacking}  \label{stacks}

Here we stack our sample in various sub-samples, to investigate relations with both local projected density and stellar mass. We present the full results in Table \ref{table:RESULTS}, but also show the results in Figures \ref{F_ion_vs_Dens_MASS}, \ref{AHa_vs_MASS} and \ref{AHa_vs_Dens}. We use median stacking throughout. Prior to stacking the spectra, we normalise them to the peak of H$\alpha$ emission. When stacking as a function of environment and stellar mass, for star-forming galaxies, we neglect AGN and sources which are too close to the AGN boundary and thus were not classified (see Table \ref{table:RESULTS}). However, note that for one case (Figure \ref{F_ion_vs_Dens_MASS}), we also evaluate the ionisation parameter including AGN.

When stacking our 92 star-forming galaxies, we find a median dust extinction of A$_{\rm H\alpha}=1.04\pm0.08$. Detailed results are presented in Table \ref{table:RESULTS}. We note that when stacking our spectra of star-forming galaxies (and also when stacking our full sample), we also reveal other lines, such as HeI5876 and H$\gamma$.

The following Sections present the results of stacking as a function of stellar mass and environmental density.

\subsubsection{Ionisation parameter} \label{ionisation_param_stack}

Our results are presented in Figure \ref{F_ion_vs_Dens_MASS}. We find that while the ionisation parameter seems to be very low at the lowest environmental densities (where we find the lowest star-formation rates), from $\log_{10}\Sigma_{10}>1.6$ we find a decrease of the ionisation parameter with increasing environmental density. We find the same qualitative behaviour when using either all sources or just star-forming galaxies (rejecting the AGNs leads to an overall ionisation parameter at essentially all environments).

Figure \ref{F_ion_vs_Dens_MASS} (and Table \ref{table:RESULTS}) also shows the dependence of the median ionisation parameter as a function of stellar mass, for our H$\alpha$ selected sources in and around the A851 cluster. We find that our lowest mass H$\alpha$ emitters have by far the highest ionisation parameters, which is consistent with e.g. \cite{Darvish15}. Higher stellar mass sources have lower ionisation parameters, but there is evidence that our high mass H$\alpha$ emitters have higher ionisation parameters than those with a ``typical" stellar mass of around $10^{10}$\,M$_{\odot}$. We note that our sample is H$\alpha$ selected, and not stellar mass selected, and that we find a strong monotonic increase of SFR with stellar mass in our stacks. Thus, for higher stellar mass galaxies, our H$\alpha$ selection only picks up the actively star-forming systems, that will likely have higher ionisation potentials due to their strongly star-forming (or AGN) nature.

\subsubsection{Balmer decrement: stellar mass dominates but environment matters} \label{balmer_indi}

\begin{figure}
  \centering
 \includegraphics[width=8.2cm]{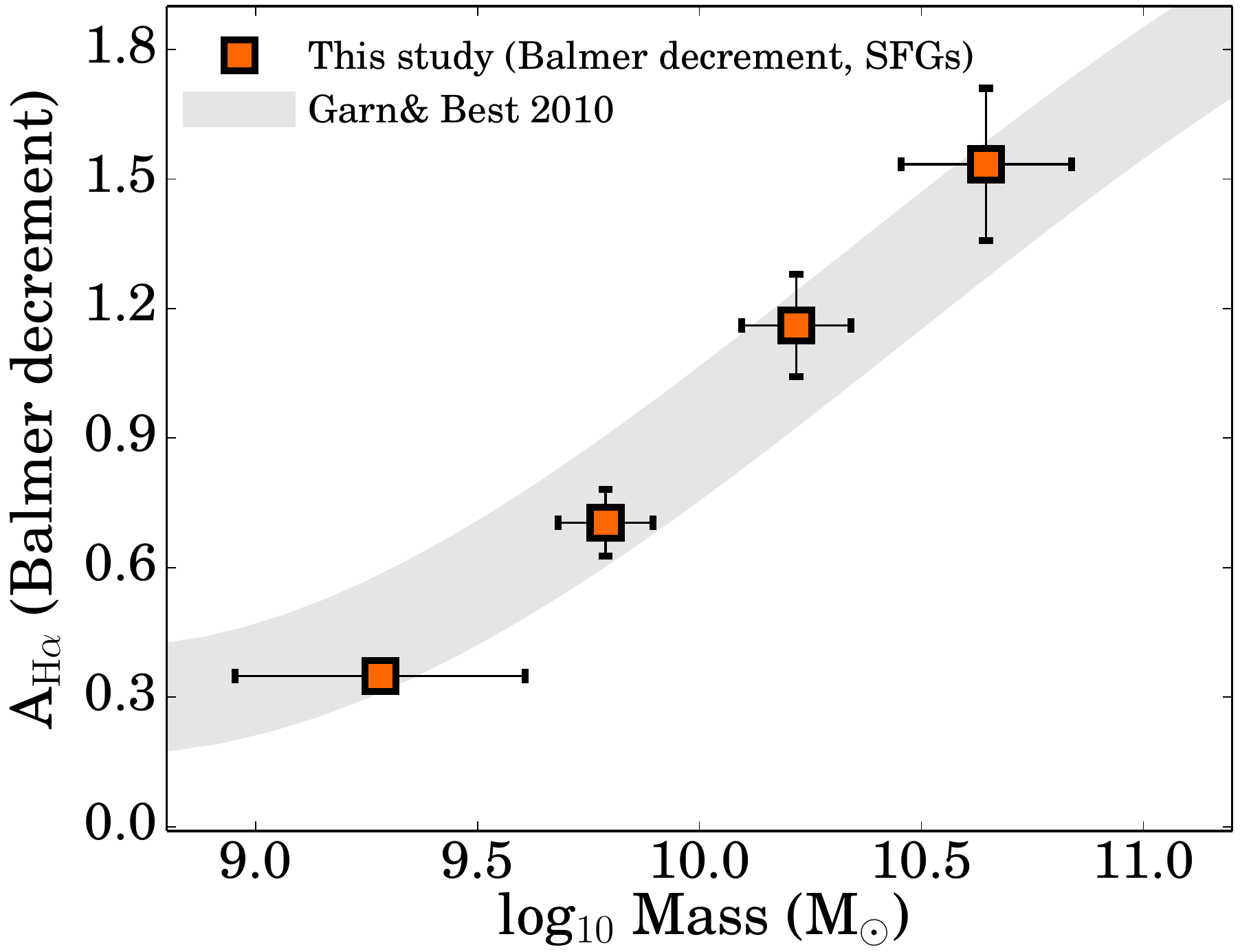}
\caption{Dust extinction (A$_{\rm H\alpha}$) obtained from the Balmer decrement (H$\beta$/H$\alpha$) from stacks, as a function of stellar mass. We find that A$_{\rm H\alpha}$ correlates strongly with stellar mass. We also show the \citealt{Garn2010} statistical relation between dust extinction and stellar mass: our results are in excellent agreement.}
\label{AHa_vs_MASS}
\end{figure}

\begin{figure}
  \centering
 \includegraphics[width=8.2cm]{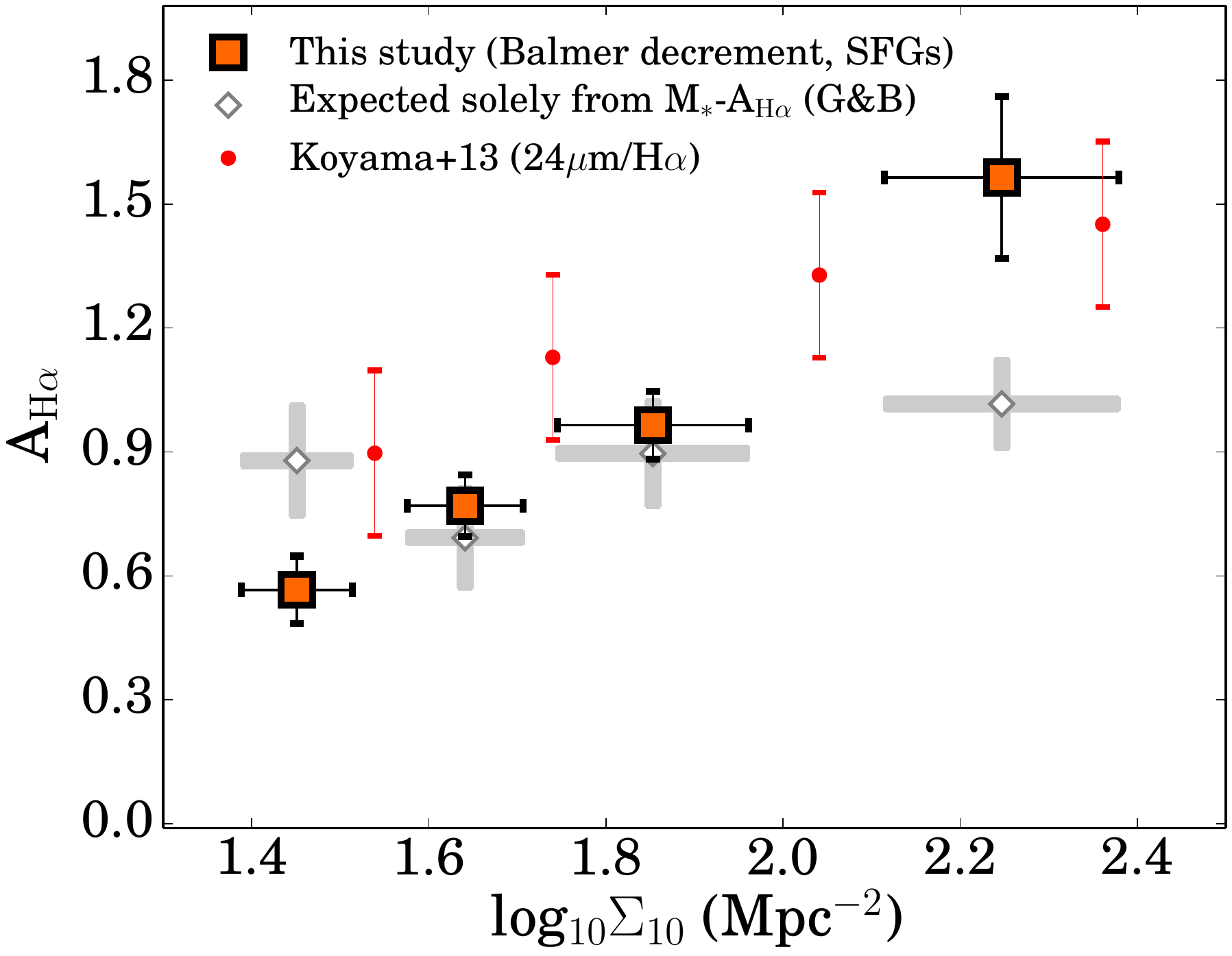}
\caption{Dust extinction (A$_{\rm H\alpha}$) obtained from the Balmer decrement (H$\beta$/H$\alpha$) as a function of local environmental density, $\Sigma$. We find that the A$_{\rm H\alpha}$ increases from the lowest to the higher density regions, in line with what was inferred from increasing 24\,$\mu$m/H$\alpha$ ratios as a function of density from \citet{Koyama13}, which we also show, for comparison. We note that A$_{\rm H\alpha}$ has been computed only for non-AGN galaxies. We note that dust extinction correlates with stellar mass in our sample, in line with \citealt{Garn2010} -- see Figure \ref{AHa_vs_MASS}. Thus, in order to take that effect into account, we use the median stellar mass of galaxies in each density bin to predict the dust extinction in that bin, which we also show. The result are consistent with those presented in \citet{Koyama13}, i.e., that the increase in dust extinction seems to be driven by environment, not mass.}
\label{AHa_vs_Dens}
\end{figure}

We evaluate the typical dust extinction (A$_{\rm H\alpha}$) of star-forming galaxies from Balmer decrements in and around the cluster as a function of both stellar mass and environmental density. The results are presented in Table \ref{table:RESULTS} and in Figures \ref{AHa_vs_MASS} and \ref{AHa_vs_Dens}.

We find that A$_{\rm H\alpha}$ increases very clearly with increasing stellar mass (Figure \ref{AHa_vs_MASS}), from A$_{\rm H\alpha}\sim0.3$ to A$_{\rm H\alpha}\sim1.6$ from the lowest to the highest stellar masses probed in our sample. Our results are in very good agreement with the relation found by \cite{Garn2010}, which shows no evolution up to at least $z\sim1-1.5$ \citep[e.g.][]{Sobral.12,Ibar.13}, and reveals that, statistically, stellar mass is a robust predictor of A$_{\rm H\alpha}$.

%
%
%
%

Nevertheless, \cite{Koyama13} found that the relation provided by \cite{Garn2010} to correct for dust extinction could not reproduce dust extinction corrections implied from the ratio 24\,$\mu$m/H$\alpha$ as a function of environmental density. In order to investigate this, we evaluate A$_{\rm H\alpha}$ as a function of environmental density in Figure \ref{AHa_vs_Dens}. Our results show that A$_{\rm H\alpha}$ rises from $0.6\pm0.2$ at the lowest densities, to A$_{\rm H\alpha}=1.6\pm0.2$ at the highest densities. Our results in Figure \ref{AHa_vs_Dens} are compared with those in \cite{Koyama13}, using 24\,$\mu$m. While the trend that we see is similar, we still find a lower normalisation, but an apparently steeper slope of the typical dust extinction as a function of environmental density. This could be a consequence of the 24\,$\mu$m results being affected by some AGN contamination (which can have significant emission of hot dust). A remaining possibility is the relation used to transform 24\,$\mu$m/H$\alpha$ ratios in A$_{\rm H\alpha}$, which is much more uncertain than Balmer decrement measurements such as the ones we obtain. 

In order to further test whether the relation we see could be driven by dust extinction correlating strongly with stellar mass, we also show in Figure \ref{AHa_vs_Dens} the dust extinction expected from the relation between stellar mass and dust extinction. We compute such dust extinctions by using the median and range of stellar masses present in each bin of local environmental density. Our results show that the relation between typical dust extinction and environmental density can not be explained by stellar mass, as expected from e.g. Figure \ref{Density-mass}.

%
%
%

\section{Conclusions} \label{Conclusion}

We have conducted a spectroscopic survey over a well-defined sample of H$\alpha$ emitters, selected with wide-field narrow-band imaging over A851/Cl 0939+4713 at $z\sim0.4$ \citep{Koyama11,Koyama13}. We measured [O{\sc ii}], H$\beta$, [O{\sc iii}], H$\alpha$ and [N{\sc ii}] for a sample of 119 H$\alpha$ emitters in and around the A851 cluster. We also use stacking and obtain Balmer decrements, metallicities and ionisation parameters to investigate if the nature, dust properties, metallicities and other properties vary as a function of environment. Our results show that:

\begin{itemize}

\item  About $70\pm13$\% of H$\alpha$ emitters in and around Abell 851 are clearly star-forming, while $17\pm5$\% are AGN. We do not find any strong dependence of the AGN fraction on environment. Thus, the rise of the typical 24\,$\mu$m/H$\alpha$ found in \cite{Koyama13} is not caused by a potential rise in the AGN fraction.

\item We find a strong mass-metallicity at all environments, with no significant dependence on environment.

\item The ionisation parameter is found to be the highest (log$_{10}$(q$_{ion}$)$\sim7.5-7.6$) for our lowest mass H$\alpha$ emitters. While we find that the ionisation parameter falls with increasing stellar masses up to $\sim10^{10}$\,M$_{\odot}$, we find that, for an H$\alpha$ selected sample, the ionisation parameter starts to increase slightly again (by $\sim0.15$\,dex) with stellar mass up to 10$^{11}$. This is likely a consequence of our selection, which yields active massive galaxies.

\item The inclusion of AGN rises the ionisation parameter at all stellar masses and at all environments.

\item We find that H$\alpha$ emitters residing in the lowest densities have the lowest ionisation parameters, and that intermediate environments show H$\alpha$ emitters with the highest ionisation parameters ($\approx7.3-7.4$). For higher environmental densities we find that the ionisation parameter declines by 0.15\,dex. Intermediate density environments show the highest [O{\sc iii}]/H$\alpha$ and [O{\sc iii}]/[O{\sc ii}] line ratios, typically twice as large as in the highest and lowest densities.

\item We find that dust extinction (A$_{\rm H\alpha}$) correlates strongly with stellar mass, as in \cite{Garn2010}, with stellar mass being the strongest predictor of A$_{\rm H\alpha}$. However, we find that our H$\alpha$ emitters at the highest and lowest environmental densities deviate significantly from what would be predicted from their stellar masses, likely hinting that the environment still plays a key role which seems independent of stellar mass.

\item Star-forming galaxies in the densest environments are found to be significantly dustier (A$_{\rm H\alpha}\approx1.5-1.6$) than those residing in the lowest density environments (A$_{\rm H\alpha}\approx0.6$). The correlation between A$_{\rm H\alpha}$ and environment is not driven by stellar mass, and deviates the most from what one would predict based on stellar mass at the lowest and the highest environmental densities, where likely more extreme processes may be happening.

\end{itemize}

\section*{Acknowledgements}

The authors thank the anonymous reviewer for many helpful comments and suggestions which improved this work. DS acknowledges financial support from the Netherlands Organisation for Scientific research (NWO) through a Veni fellowship and from FCT through a FCT Investigator Starting Grant and Start-up Grant (IF/01154/2012/CP0189/CT0010). AS acknowledge financial support from an NWO top subsidy (614.001.006). B.D. acknowledges financial support from NASA through the Astrophysics Data Analysis Program (ADAP), grant number NNX12AE20G. This research has made use of NASA's Astrophysics Data System. The authors acknowledge the award of time (W14BN020) on the William Herschel Telescope (WHT). WHT and its service programme are operated on the island of La Palma by the Isaac Newton Group in the Spanish Observatorio del Roque de los Muchachos of the Instituto de Astrofisica de Canarias.

\bibliographystyle{mnras}
\bibliography{myBib.bib}

\appendix

\section{Catalogue of H$\alpha$ emitters}

Here we present the full catalogue of spectroscopically confirmed H$\alpha$ emitters in and around the cluster.

%
%
%
%
\begin{table*}
\begin{center}
{\scriptsize
\begin{tabular}{lccccccccccccc}
\hline
\noalign{\smallskip}
ID    & $\alpha_{\rm J2000}$ & $\delta_{\rm J2000}$   & $z_{spec}$ & z$_{\rm AB}$ & F$_{\rm H\alpha}$ & $\log\,\Sigma_{10}$    & [N{\sc ii}]/H$\alpha$     & [O{\sc iii}]/H$\beta$     &  [O{\sc iii}]/H$\alpha$ & [O{\sc ii}]/H$\alpha$   & Mass  & Flag  & AGN \\
            &               &                   &        &            &      log$_{10}$         &            &      &  &        &      & log$_{10}$(M$_{\odot}$)  &   &            \\
\hline
  DUSQ\_135 & 09:41:29.614 & +47:00:42.12 & 0.4031 & 22.3 & -15.8 & 1.91 & 0.27 & 0.78 & 0.12 & 0.11 & 9.41 & 3 & 0\\
  DUSQ\_164 & 09:44:02.183 & +46:49:23.80 & 0.3887 & 22.0 & -15.7 & 1.51 & 0.28 & 2.67 & 1.44 & 2.18 & 9.52 & 1 & 0\\
  DUSQ\_202 & 09:42:01.043 & +46:49:02.72 & 0.3976 & 22.2 & -15.8 & 1.45 &  &  & 0.73 & 0.70 & 9.93 & 3 & -99\\
  DUSQ\_207 & 09:43:54.517 & +46:49:14.63 & 0.4050 & 22.6 & -15.9 & 1.59 & 1.14 & 0.57 & 0.14 & 0.12 & 9.72 & 1 & 1\\
  DUSQ\_208 & 09:44:04.349 & +46:50:10.49 & 0.3891 & 21.7 & -15.6 & 1.64 & 0.25 & 0.80 & 0.23 & 0.19 & 9.70 & 3 & 0\\
  DUSQ\_220 & 09:43:11.510 & +46:45:26.47 & 0.4104 & 23.0 & -15.8 & 1.37 & 0.32 & 1.69 & 0.22 & 0.14 & 9.46 & 3 & 0\\
  DUSQ\_283 & 09:43:26.618 & +47:06:59.78 & 0.4036 & 21.1 & -15.4 & 1.82 & 0.27 & 0.66 & 0.31 & 0.09 & 10.21 & 3 & 0\\
  DUSQ\_305 & 09:41:37.889 & +46:54:10.45 & 0.4051 & 22.9 & -15.7 & 1.50 & 0.50 & 0.33 & 0.18 & 0.16 & 9.31 & 3 & 0\\
  DUSQ\_307 & 09:43:38.247 & +46:54:38.57 & 0.3885 & 23.5 & -15.8 & 1.81 & 0.12 & 8.40 & 1.74 & 0.67 & 8.70 & 1 & 2\\
  DUSQ\_320 & 09:43:54.469 & +46:57:15.19 & 0.4059 & 22.7 & -15.7 & 1.52 &  & 3.59 & 1.13 & 1.55 & 9.35 & 3 & -1\\
  DUSQ\_323 & 09:42:38.664 & +46:58:02.68 & 0.4097 & 23.0 & -15.7 & 1.94 & 0.17 & 0.28 & 0.07 & 0.21 & 9.12 & 3 & 0\\
\hline
\end{tabular}
}
\caption{H$\alpha$ emitters in our spectroscopic sample of 119 sources with Flags 1,2,3. We show 11 example entries: the full catalogue is published in the on-line version of the paper. AGN flags: -99 and -1: Unclass; 0: SFGs; 1: Others; 2: AGN.}\label{table:Gal_props}
\end{center}
\end{table*}

\bsp
\label{lastpage}
\end{document}